\documentclass[twocolumn,noshowpacs,preprintnumbers,amsmath,amssymb]{revtex4}

\usepackage{verbatim}

\usepackage{graphicx}
\usepackage{dcolumn}
\usepackage{bm}
\usepackage{wrapfig}

\begin{document}

\title{On the Electric Conductivity of Highly Ordered Monolayers of Monodisperse Metal Nanoparticles}

\author{Denis Greshnykh}
\author{Andreas Fromsdorf}
\author{Horst Weller}
\author{Christian Klinke}
\email{klinke@chemie.uni-hamburg.de}
\affiliation{Institute of Physical Chemistry, University of Hamburg, Germany}

\begin{abstract} % abstract goes here

Monolayers of colloidally synthesized cobalt-platinum nanoparticles of different diameters characterized by TEM (transmission electron microscopy) were deposited on structured silicon oxide substrates and characterized by SEM (scanning electron microscopy), GISAXS (grazing incidence x-ray scattering), and electric transport measurements. The highly ordered nanoparticle films show a thermally activated electron hopping between spatially adjacent particles at room temperature and Coulomb blockade at low temperatures. We present a novel approach to experimentally determine the particles charging energies giving values of 6.7-25.4~meV dependent on the particles size and independent of the interparticle distance.
These observations are supported by FEM (finite element method) calculations showing the self-capacitance to be the determining value which only depends on the permittivity constant of the surrounding space and the particles radius.

\end{abstract}

\maketitle

Nanostructured materials have raised substantial interest in research and industry \cite{Gleiter,Moriarty,Rao,Feldheim,Rittner}. They are used or discussed for a wide range of applications \cite{Shipway}, like in optical devices \cite{Schmid}, or in electronic \cite{Likharev} and magnetic architectures \cite{Terris,Lu}. For the chemical industry nanostructured catalysts are of tremendous interest \cite{Lu,Toshima,Mo,Hu}. Also under intensive research are applications in sensors \cite{Vossmeyer,Arai,Yonzon} and medicine \cite{Lu,Gleich}. Furthermore, nanoparticles and assemblies of nanoparticles are interesting model systems for investigations of the matter properties on the nanometer scale \cite{Beverly,Markovich}. Size and form dependence of physical and chemical properties and photophysical processes in constrained systems are still under investigation \cite{NanoscaleMaterialsInChemistry,El-Sayed}. While the building blocks of conventional macroscopic crystals are atoms or molecules, nanoparticles can compose ``artificial solids'' or ``supercrystals'' \cite{Kastner,Beverly}.
The use of nanoparticles as building blocks allows control over certain properties of these novel materials.
The electrical transport properties of these systems are the subject of numerous publications \cite{Feldheim,Schmid,Likharev,LB-Paper,Simon,Quinn,Sampaio}. Ordered monolayers of metal nanoparticles represent a simple system which allows verification of theoretical models.
A vast number of experimental and theoretical results dealing with the transport properties of the nanoparticle arrays has been published \cite{Black,HoppingReview,MurrayTransportmodel,ClassicCoulombBlockade,Peng,HeathRemacleLevineSurfPot,WellerElektro,MiddeltonWingreen,ActivatedRegime,WellerTransportInSmall,TobiasFaradayDiscuss,Simon} (to mention only a few). Despite this tremendous work many features of this systems are still unclear.

\begin{figure}[htbp]
  \centering
  \includegraphics[width=0.49\textwidth]{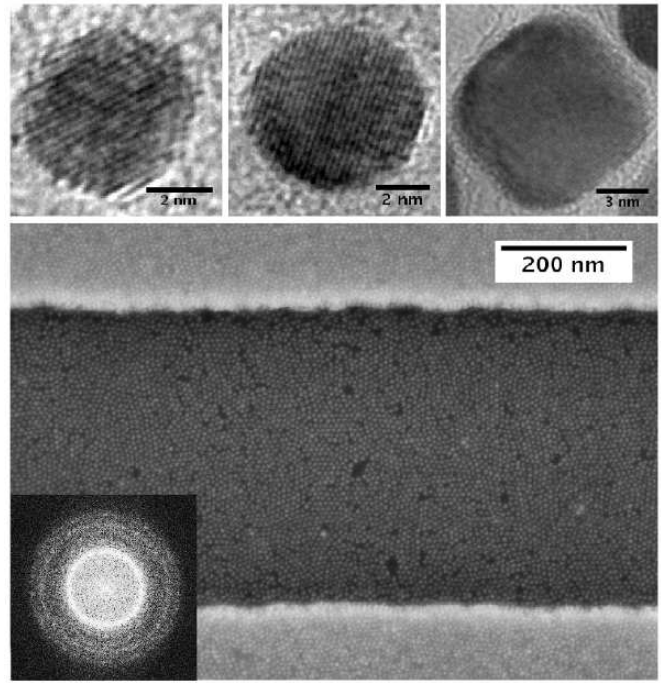}
  \caption{High resolution TEM images of the monodisperse cobalt-platinum nanoparticles used in this work (upper figures). The big particles were cubic in contrast to the other nanocrystals. SEM micrograph of a cobalt-platinum nanoparticle monolayer deposited on an substrate structured with gold electrodes (in top and bottom area of the lower picture). The inset shows the obtained Fourier transformed pattern.}
  \label{fig:TEMs}
\end{figure}

In this paper we discuss the different transport properties of nanoparticle films and supercrystals which are governed in contrast to conventional crystals or metal films by the presence of tunnel barriers between the particles if the tunneling resistance ($R_t$) considerably exceeds the resistance quantum $R_t \gg \frac{h}{e^2}= 25.8 k\Omega$, with the Planck constant $h$ and the elementary charge $e$. At low temperatures small nanoparticles show current suppression known as the Coulomb blockade \cite{CoulombBlockade} which  originates from a charging energy $E_{\text{c}}$ of the particles due to their very small capacitance. The total capacitance of the particles $C_{\Sigma}$ is the sum of the self-capacitance $C_s$ and the contact capacitance of a particle to the adjacent particles $C_{\text{geo}}$.
\begin{align}
  & E_{\text{c}} = \frac{e^2}{2C_{\Sigma}} \qquad \text{with} \qquad C_{\Sigma} = C_s + C_{\text{geo}}
  \label{eq:CoulombBlockade}
\end{align}
The observation of the Coulomb blockade requires the thermal energy to be lower than the particles charging energy: $E_{\text{c}} \gg k_{\text{B}}T$.

\begin{figure}[htbp]
  \centering
  \includegraphics[width=0.45\textwidth]{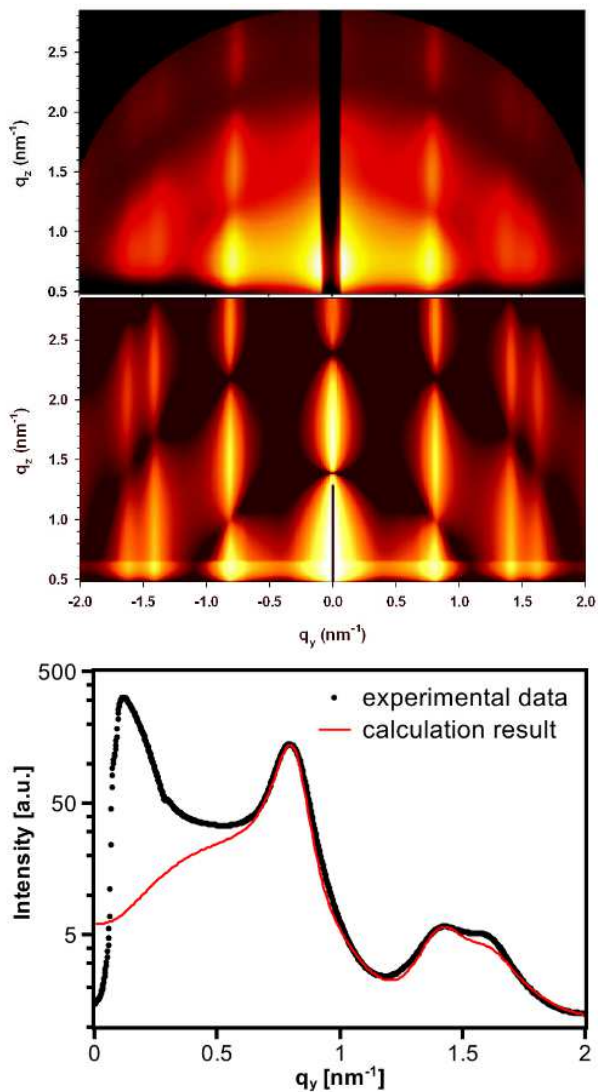}
  \caption{GISAXS pattern and IsGISAXS simulation of cobalt-platinum particles on silicon substrate. Simulation model: 2D hexagonal lattice (nearest neighbor distance = 8.8\,nm) of flattened spheroidals on a substrate with lateral diameters of 7.6\,nm and a height of 6.5\,nm. Lower graph: GISAXS out of plane scattering curves (at Yoneda maximum) from cobalt-platinum particles on silicon. Curves were analyzed using a 2D hexagonal lattice function ($ q = 8.8 nm$, spherical form factor with diameter of 7.6\,nm). The calculation of the GISAXS out of plane cuts along $q_y$ is based on a model
which does not take uncorrelated roughness and diffuse scattering into account. Directed to low $q$ these properties lead to an increasing intensity.}
  \label{fig:GISAXSresults}
\end{figure}

A bottom-up approach was used to synthesize cobalt-platinum nanoparticles by wet chemical methods formerly developed in our group \cite{SynthesisPaper}. We managed to synthesize particles of different diameters by injecting the cobalt precursor solution at appropriate temperatures of the platinum precursor/ligands solution. The resulting nanocrystals showed a chemically disordered fcc lattice as was shown by XRD and electron diffraction. The composition of Co$_{0.14-0.22}$Pt$_{0.86-0.78}$ for different syntheses was determined by EDX (energy dispersive x-ray analysis). The particle diameters were investigated by TEM, the obtained values were 4.4\,nm, 7.6\,nm, and 10.4\,nm (\ref{fig:TEMs}).

\begin{figure}[htbp]
  \centering
  \includegraphics[width=0.45\textwidth]{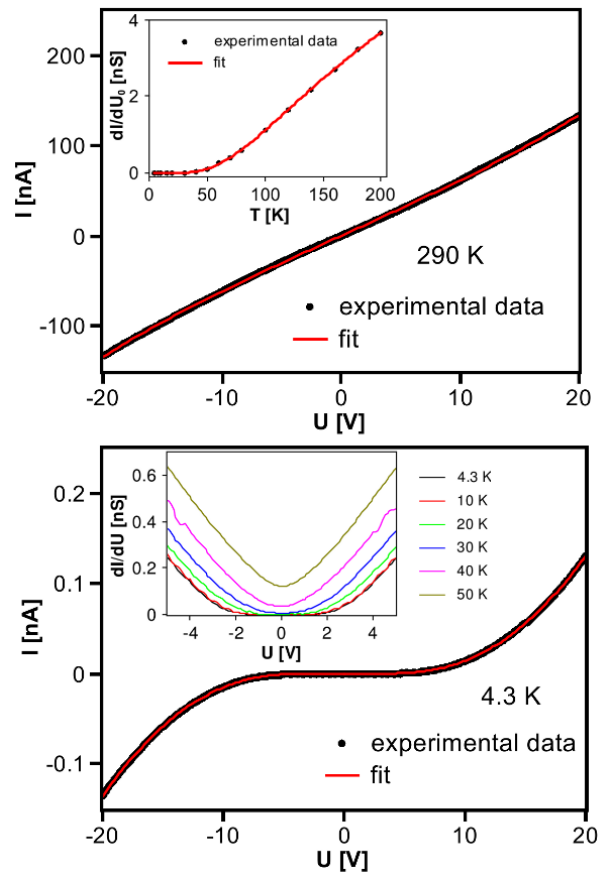}
  \caption{As an example the I-V curves of the 7.6\,nm particles at different temperatures are shown. The films consisting of other particle sizes showed qualitatively the same behavior. The I-V curves showed distinct deviations at different temperatures. While the curve at room temperature is almost linear (upper graph) a clear current blockade is present at 4.3\,K (lower graph). While at room temperature our data followed a $\sinh$ function, the Nordheim-Fowler equation was applicable at low temperatures. The inset at the upper graph shows the development of the experimental zero bias conductance of the same film with temperature. The corresponding fit was calculated using the simple thermally activated model. The inset of the lower figure shows the evolution of the differential conductance with temperature for the 7.6\,nm particles.}
  \label{fig:IV}
\end{figure}

For GISAXS and electrical measurements we deposited the nanoparticles from the toluene solution on silicon oxide substrates by the Langmuir-Blodgett technique as we have demonstrated recently \cite{LB-Paper}. SEM micrographs showed superlattices of hexagonally ordered monolayers with a few small cavities (see \ref{fig:TEMs}).

The microscopic methods SEM and AFM (atomic force microscopy) are not suitable to investigate the surface morphology on the scale of several millimeters. Micrographs show only a small area of the sample and for better statistics additional methods are needed. Here, GISAXS is the first choice, as it reveals information from a large surface area (several mm$^2$) of the surface. The scattering curves can be compared with the Fourier transformations of the microscopic images, to demonstrate that the same nanostructure is uniformly spread over a large surface area. From simulations of GISAXS patterns it is possible to get information about the form factor and the interference function, which leads to the size and shape of the particles and their lateral distances \cite{Froemsdorf}.

\begin{figure}[htbp]
  \centering
  \includegraphics[width=0.45\textwidth]{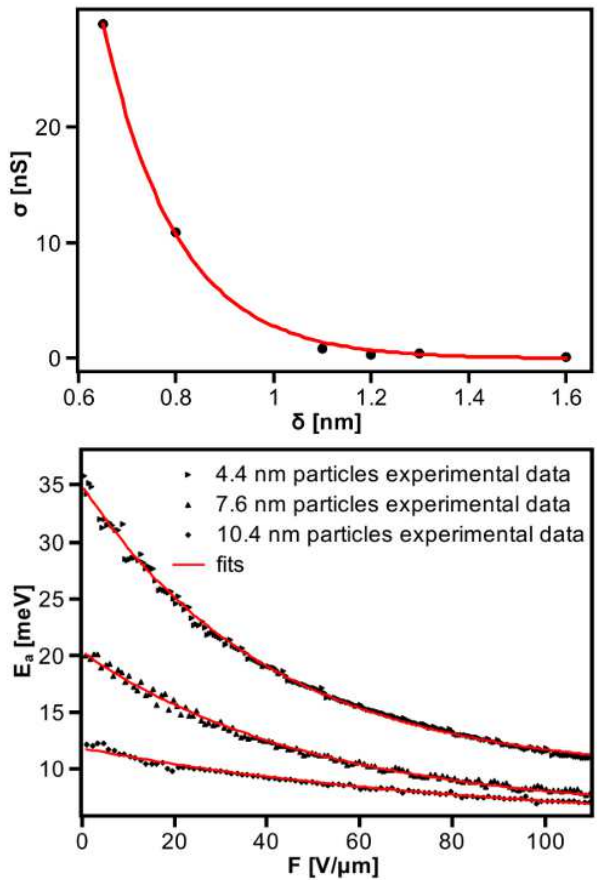}
  \caption{The determined film conductivity scaled exponentially with the interparticle distance (upper figure). Field dependence of the activation energy was well-fit by an exponentially decaying equation (lower graph).}
  \label{fig:SigmaDecay}
\end{figure}

The measured GISAXS patterns can be compared with simulations done by the IsGISAXS software \cite{IsGISAXS}. \ref{fig:GISAXSresults} shows such a pattern and a simulation for cobalt-platinum nanoparticle monolayer. Assuming the depicted model of flattened spheroidal particles packed in a two-dimensional array the simulation leads to an almost identical pattern. The dark line indicates were the detector was protected with a lead stripe to prevent damage due to high scattering intensities. This method allows determining the lateral distances between the particles and their height and shape, which are identical with the calculations made by analyzing the intensity cut along $q_y$ at the critical angle of the substrate \cite{Froemsdorf2}. This analysis is shown in \ref{fig:GISAXSresults}. GISAXS allowed us to determine not only the long range ordering in our films but also to determine precisely the shortest surface to surface interparticle distances.

In order to probe the transport properties of the nanoparticle films we measured the DC response at different temperatures.
Although the big particles were cubic no form induced effects were detected in our electrical measurements.
At room temperature the obtained I-V curves are point-symmetric and almost linear. Toward lower temperatures, however, the current-voltage characteristic becomes more and more nonlinear and the device is clearly in Coulomb blockade regime at moderate voltages as can be seen in \ref{fig:IV}. Due to the nonlinearity of the I-V curves at any temperature and the smooth transition from the blocked regime into unblocked with rising voltage it is not appropriate to extract the blockade voltage using linear interpolation. Also the transition from noise level to measurable current levels was not applicable to determine the Coulomb blockade as the blockade voltage determined this way was dependent on the measurements temperature and can be expected to depend on the noise level of the used experimental setup.

\begin{figure}[htbp]
  \centering
  \includegraphics[width=0.45\textwidth]{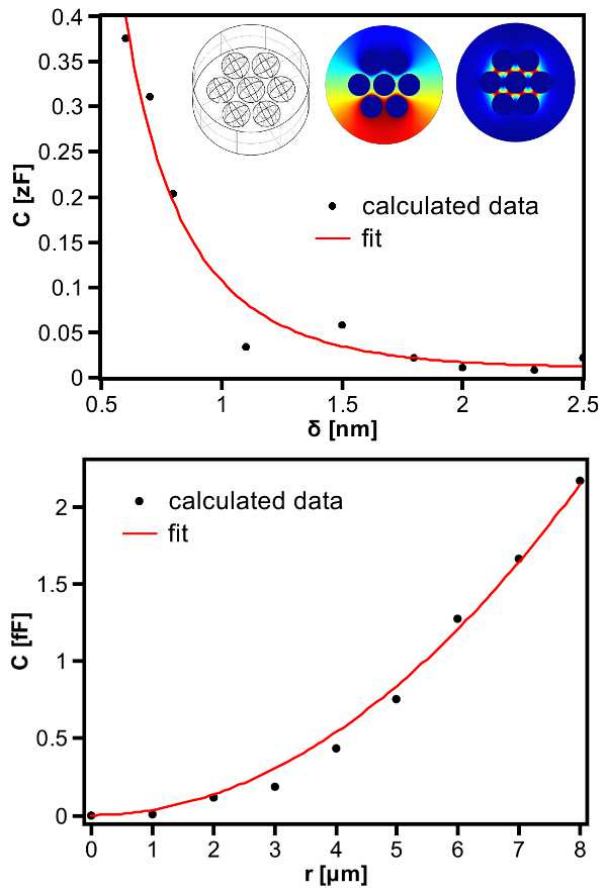}
  \caption{Calculated capacitance values of a sphere in a hexagonal lattice for different spheres radii and intersphere distances with corresponding numerical fits. The inset of the upper figure shows the geometry used for FEM calculations, the horizontal sections of the calculated potential distribution, and the normalized electric field.}
  \label{fig:FEM}
\end{figure}

At room temperature the I-V curves could be fitted well by a $\sinh$ function suggested in literature for electron hopping between spatially adjacent energy states in semiconductors \cite{Singh}. In contrast at 4.3\,K the I-V curves can be fitted by the Nordheim-Fowler equation typical for tunnel emission, indicating a different transport mechanism at low temperatures (\ref{fig:IV}).
Simple exponential fits and fits by power law were not appropriate to describe our data.
From the measurements at different temperatures we can deduce a negative temperature coefficient of resistivity which is indicative of thermally activated transport. Examining the differential conductance (\ref{fig:IV}) gives a deeper insight into the charge transport mechanism.
For $T< 40 K$ the electrons are tunneling when the applied field strength is sufficient to overcome the particles charging energy. For $T\geq\ 40 K$ the electrons are thermally excited and thermally activated hopping can be expected. In the literature different hopping mechanisms have been proposed: the simple thermally activated hopping \cite{NeugebauerWebb} and variable range hopping (VRH) suggested for granular metals which are typical highly disordered systems \cite{AbelesModell}. For thermally activated transport the conductivity $\sigma$ scales with
\begin{equation}
  \sigma \propto \exp{\left( -\frac{E_{\text{a}}}{k_{\text{B}}T} \right)}.
\end{equation}
Here $E_{\text{a}}$ is the activation energy and $k_{\text{B}}$ is the Boltzmann constant.
For the VRH driven transport a different correlation was proposed \cite{AbelesModell}:
\begin{equation}
  \sigma \propto \exp{\left( -\frac{T_0}{T} \right)^{\frac{1}{2}}}.
\end{equation}
Here $T_0$ can be interpreted as the activation temperature.
While the first mechanism describes the charge transfer between spatially adjacent particles, the VRH mechanism considers the transfer between the energetically closest particles states. In order to distinguish between these two cases we plotted the zero conductance data in an appropriate way for temperatures where the Coulomb blockade was overcome (see Supporting Information). From the linear fits we conclude that the simple thermally activated transport model describes our data better than any other applied models.
From the slope of the Arrhenius plot a value of 20\,meV for the activation energy was obtained for the 7.6\,nm particles. For smaller particles a value of 36\,meV was obtained while for the 10.4\,nm particles a value of 12\,meV was determined. In the literature \cite{SchmidSimon} this activation energy was identified as the particles charging energy introduced in Equation \ref{eq:CoulombBlockade}; however, the Coulomb charging energy from Equation~\ref{eq:CoulombBlockade} should depend linearly on the particles' array width along the charge transport direction.
Our data suggests instead that the activation energy scales with the particles' size. Small particles possess lower self-capacitance, which results in higher activation energy according to Equation~\ref{eq:CoulombBlockade}.
In contrast, varying the interparticle distance did not affect the activation energy; our measurements always gave values in the range of 18--20\,meV for the 7.6\,nm particles for the range of particle separations studied.
To vary the interparticle distance we replaced the hexadecylamine ligands present after synthesis by dodecylamine, tetradecylamine and octadecylamine. For these cases, the interparticle distances as determined by GISAXS varied between 0.7\,nm and 1.6\,nm.
At room temperature the film conductivity was found to scale exponentially with the interparticle spacing (see \ref{fig:SigmaDecay}).

The striking feature of the obtained data is the nonlinearity of the I-V curves even at higher voltages (see \ref{fig:IV}).
This observation encouraged us to verify the voltage dependence of the activation energy. In contrast to the results published so far we observed exponential decay of the activation energy at rising field strength (\ref{fig:SigmaDecay}).
In fact, we find that the activation energy is not equal the charging energy. At low fields electrons have to overcome the tunneling barrier and the particles' charging energy. Rising field strength increases the electron energy until the Fermi energy of the electron sea in particle matches the energy of empty states of a nearby particle, into which the electrons are injected. In this regime the charging energy $E_{\text{c}}$ is overcome by tunneling electrons, thermally assisted tunneling is not affected by further increase in voltage. From this point of view the particles charing energy can be calculated as the difference between the experimental zero field activation energy $E_0$ and the high field limit activation energy $E_{\text{hf}}$. Then the measured activation energy can be expressed by the following equation.
\begin{align}
   E_{\text{a}} & = E_{\text{hf}} + \left( E_0 - E_{\text{hf}} \right) \exp{\left( -\frac{U}{U_0} \right)} \\
       & = E_{\text{hf}} + \left( E_{\text{c}} \right) \exp{\left( -\frac{U}{U_0} \right)}
\end{align}
Here $U_0$ represents a fitting parameter with voltage units.
For small particles the Coulomb blockade was found to be 25.4\,meV, for midsized particles 14.1\,meV, and for big particles 6.7\,meV.

Equation~\ref{eq:CoulombBlockade} suggests that one can calculate the charging energy from a knowledge of $C_{\Sigma}$, the total capacitance ``felt'' by the tunneling electrons. $C_{\Sigma}$ can be calculated from the sum of the particles' self-capacitance $C_{\text{self}}$ and the interparticle junction capacitance $C_{\text{geo}}$. The self-capacitance of a free standing sphere can be calculated analytically using equation~\ref{eq:selfcapacitance}.
\begin{equation}
  C_{\text{self}} = 4\pi\varepsilon_0\varepsilon_r r
  \label{eq:selfcapacitance}
\end{equation}
Here $\varepsilon_r$ is the relative permittivity, $\varepsilon_0$ the vacuum permittivity of the interparticle environment and $r$ is the sphere radius.
The junction capacitance depends on the geometry of the system under investigation and usually cannot be calculated analytically. In the literature the model of a metallic sphere with dielectric shell embedded in metallic continuum has been used in many cases \cite{Wuelfing,Peng,Joseph,LB-Paper} after the paper of Abeles et al. \cite{AbelesModell}. Alternatively, the capacitance of a hemisphere over a plane surface has been approximately calculated \cite{Black}. These models deviate considerably from the experimental geometry and potential distribution. In order to obtain the value of the Coulomb blockade we computed the geometrical capacitance by means of the finite element method (FEM). As the model a hexagonal lattice of spheres with 7.6\,nm diameter was used (\ref{fig:FEM}). This diameter corresponds to the medium-sized nanoparticles used in our experiments. A linear potential gradient was applied to the spheres. For the underlying oxide layer a dielectric constant $\varepsilon_r = 4.48$ was used \cite{HCP} which had only a slight influence on the obtained values. For the spheres environment the vacuum permittivity was used. The corresponding capacitance values for diverse dielectric environments can be calculated easily as the capacitance is proportional to the dielectric constant. Also a model with $30^{\circ{}}$ rotated potential gradient was used giving negligible deviations. The calculations gave a very low capacitance of $1.1 \cdot 10^{-23}]{F}$. This result shows that in contrast to the results reported previously using less appropriate models the tunneling junction capacitance can be neglected compared to the self-capacitance of particles as was assumed previously by Averin and Korotkov in \cite{Averin}. Nevertheless it is not possible to obtain the activation energy from the calculated capacitance as the value of the dielectric constant of the interparticle environment in the experimental setup is not known.

In order to verify at which point the geometrical capacitance starts to dominate over the self-capacitance, we calculated geometrical capacitances for different particle sizes and physically reasonable interparticle distances. The resulting values of capacitance versus radius can be fit well by a parabolic equation, while the values obtained from the interparticle distance variation were fitted by a polynomial equation:
\begin{equation}
  C(\delta) = a + b\delta^{-1} + c\delta^{-2}.
  \label{eq:FEMfit}
\end{equation}

In equation~\ref{eq:FEMfit} $a$, $b$, and $c$ are fit parameters, and $\delta$ is the shortest distance between the particles' surfaces.
From the calculated fit parameters we were able to obtain a universal equation which allows one to calculate the capacitance of a sphere in an hexagonal lattice with various sphere sizes and intersphere distances. If the relative dielectric constant of the interparticle environment is known, the calculated capacitance has to be multiplied  additionally by this value.
\begin{align}
  & C_{\text{geo}}(r,\delta) = 6.25\cdot10^{16}m^{-2} \varepsilon_r \left( a + b\delta^{-1} + c\delta^{-2} \right) r^2 \\
  & a 			= 8.6\cdot 10^{-24}~F		\\
  & b 			= -2.5\cdot 10^{-32}~Fm		\\
  & c 			= 1.5\cdot 10^{-40}~Fm^{2}
\end{align}
For interparticle distances smaller than 1\,nm, the spheres diameter has to exceed approximately $1 \mu m$ in order for the geometric capacitance to overcome the particles self-capacitance.

In the present work we demonstrated the utilization of the Langmuir-Blodgett technique for deposition of cobalt-platinum nanoparticles of different sizes (4.4\,nm, 7.6\,nm and 10.4\,nm diameter) on structured silicon oxide substrates. The obtained ordered particle monolayers were characterized by SEM and GISAXS.

The DC measurements on the films at temperatures be\-tween 50 and 300\,K showed a simple thermally activated electron hopping between spatially adjacent particles with activation energies in the range of 12--36\,meV. Small particles showed the highest values and big particles the smallest. Calculations and experimental results suggest the charging energy to depend only on the particles' self-capacitance and to be independent of the interparticle distance and thus independent of the geometrical capacitance, which was shown by FEM calculations to be negligible in the nanoparticle arrays. The geometrical capacitance starts to be predominant in comparison to the self-capacitance when the particles' diameter exceeds one micrometer. Therefore the charging energy of smaller particles depends only on the dielectric constant of the interparticle medium and on the particles' radius.

The I-V curves were fit by equations derived for semiconductors, indicating the similarity between the transport mechanisms in semiconductors and metal nanoparticle arrays. Indeed the charging energy in the metal nanoparticle arrays plays an analogous role as the band gap in semiconductors.

The investigated nanoparticle films showed an exponential decrease of the activation energy with the applied electric field. This dependence allowed us to calculate the particles charging energy, giving values in the range of $6.7-25.4 meV$. To our knowledge this approach has not been previously reported.

The long range ordering of particles in our films gave rise to the observed effects which were explained by a simple next neighbor electron hopping. The size dependent effects are supported by numerical calculations.

\end{document}